\documentclass[a4paper]{article}

\usepackage{INTERSPEECH2020}
\usepackage{amsmath,graphicx,amssymb}
\usepackage[export]{adjustbox}
\usepackage[caption=false]{subfig}
\usepackage{color}
\usepackage{xcolor}
\usepackage{enumitem}
\usepackage{multirow}
\PassOptionsToPackage{hyphens}{url}\usepackage{hyperref}

\hypersetup{colorlinks=true}

\title{VoiceFilter-Lite: Streaming Targeted Voice Separation \\
for On-Device Speech Recognition}
\name{Quan Wang \quad Ignacio Lopez Moreno \quad Mert Saglam \quad Kevin Wilson \quad Alan Chiao \\
Renjie Liu \quad Yanzhang He \quad Wei Li \quad Jason Pelecanos \quad Marily Nika \quad Alexander Gruenstein}
\address{
  Google LLC, USA
}
\email{\href{mailto:quanw@google.com}{\nolinkurl{quanw@google.com}}}

\begin{document}

\maketitle
\begin{abstract}
We introduce VoiceFilter-Lite, a single-channel source separation model that runs on the device to preserve only the speech signals from a target user, as part of a streaming speech recognition system. Delivering such a model presents numerous challenges: It should improve the performance when the input signal consists of overlapped speech, and must not hurt the speech recognition performance under all other acoustic conditions. Besides, this model must be tiny, fast, and perform inference in a streaming fashion, in order to have minimal impact on CPU, memory, battery and latency. We propose novel techniques to meet these multi-faceted requirements, including using a new asymmetric loss, and adopting adaptive runtime suppression strength. We also show that such a model can be quantized as a 8-bit integer model and run in realtime.
\end{abstract}

\vspace{-3pt}
\section{Introduction}
\vspace{-3pt}
As deep learning gains its popularity in speech and signal processing, we have seen impressive progress in recent years using blind source separation techniques for multi-talker speech recognition. This is often referred to as the ``cocktail-party problem''~\cite{cherry1953some}. Some of the representative advances include deep clustering~\cite{hershey2016deep}, deep attractor network~\cite{chen2017deep}, and permutation invariant training~\cite{yu2017permutation,kolbaek2017multitalker}.

\emph{Voice filtering}, also known as \emph{speaker extraction}, is a slightly different problem than blind speech separation. In the voice filtering setup, we aim to separate the speech of a target speaker, whose identity is known. The target speaker could be represented by either a one-hot vector from a closed speaker set~\cite{zmolikova2017speaker}, or a speaker-discriminative voice embedding, such as an i-vector~\cite{dehak2010front} or d-vector~\cite{ge2e}. Several different frameworks have been developed for the task of speaker-conditioned speech separation in recent years, including DENet~\cite{wang2018deep}, SpeakerBeam~\cite{delcroix2018single}, and VoiceFilter~\cite{wang2018voicefilter}. A speaker-conditioned separation system has several advantages over blind separation systems that make it more applicable in real applications: (1) It is not required to know the number of sources, which is usually not available at runtime. The same model works for both single-speaker and multi-speaker scenarios. (2) It does not have the speaker-permutation problem. (3) It does not require an output channel selection step after the separation, which in blind separation systems is usually implemented by selecting the loudest output channel, or employing an additional speaker verification system.

However, integrating a speaker-conditioned speech separation model to production environments, especially on-device automatic speech recognition (ASR) systems such as~\cite{he2019streaming}, presents a number of challenges. Quality wise, the model should not only improve the ASR performance when there are multiple voices, but also should be harmless to the recognition performance under other scenarios, \emph{e.g.} when the speech is only from the target speaker, or when there is non-speech noise such as music present in the background. For streaming systems, to have minimal latency, bi-directional recurrent layers or temporal convolutional layers shall not be used in the model. For on-device systems, the model must be tiny and fast to require minimal budget in CPU and memory.

To achieve these goals, we implemented a new architecture which we call VoiceFilter-Lite\footnote{\footnotesize More details available at:   \url{https://google.github.io/speaker-id/publications/VoiceFilter-Lite}}. In this architecture, the voice filtering model operates as a frame-by-frame frontend signal processor to enhance the features consumed by the speech recognizer, without reconstructing audio signals from the features. The key novel contributions of this work include: (1) A system to perform speech separation directly on ASR input features; (2) An asymmetric loss function to penalize over-suppression during training, to make the model harmless under various acoustic environments (Section~\ref{sec:asym_loss}); (3) An adaptive suppression strength mechanism to  adapt to different noise conditions (Section~\ref{sec:adapt_weight}). We also provided several approaches to make the model tiny and fast to meet strict on-device production requirements, such as limiting the model topology and quantizing the model to 8-bit integer format.

\vspace{-3pt}
\section{Review of the VoiceFilter system}
\vspace{-3pt}
Our new architecture shares many common designs with our previous VoiceFilter system~\cite{wang2018voicefilter}, and here we give a brief introduction of the VoiceFilter base system.

At inference time, the VoiceFilter system takes two audio as input: noisy audio to be enhanced, and reference audio which is from the target speaker. A pre-trained speaker encoder LSTM~\cite{ge2e} is used to produce the speaker-discriminative embedding, \emph{a.k.a.} d-vector, from the reference audio. The time-frequency spectrogram of the noisy audio is computed via short-time Fourier transform (STFT), which first goes through convolutional layers, then is frame-wise concatenated with the d-vector, and finally goes through LSTM and fully connected layers to predict a time-frequency soft mask. The mask is element-wise multiplied to the spectral magnitude within the spectrogram  to reconstruct enhanced audio via inverse STFT~\cite{griffin1984signal}.

At training time, the noisy audio is generated by summing the waveforms of an interference audio and the clean audio, where the clean audio is from the same speaker as the reference audio, and the interference audio is from a different speaker. The VoiceFilter network is trained by minimizing a loss function which measures the difference between the clean spectrogram and the masked spectrogram.

\begin{figure*}[t!]
    \subfloat[\label{fig:VoiceFilterLite_IO}]{%
       \includegraphics[width=0.3\textwidth]{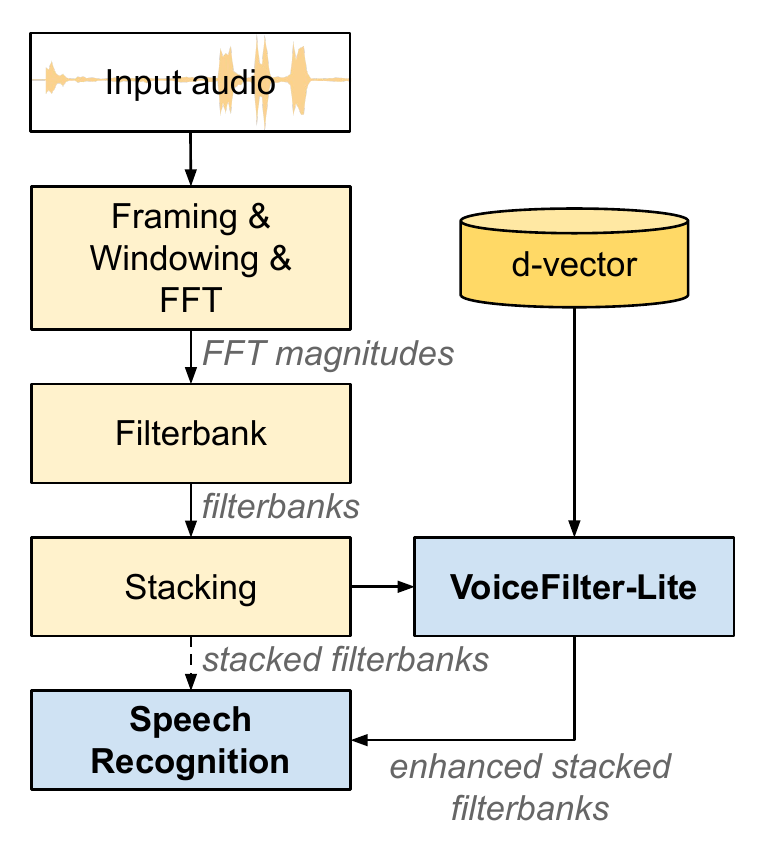}
     }
     \subfloat[\label{fig:topology}]{%
       \includegraphics[width=0.69\textwidth]{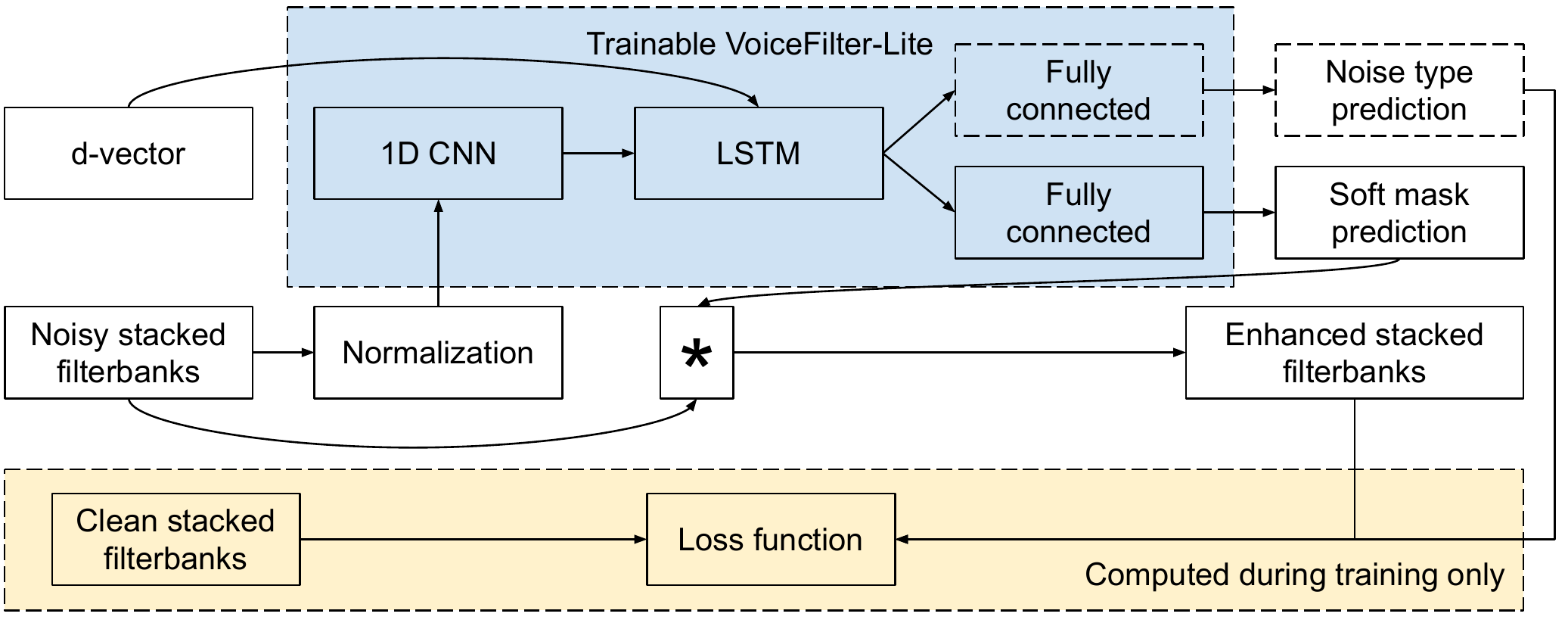}
     }
     \vspace{-5pt}
     \caption{VoiceFilter-Lite architecture, assuming using stacked filterbank energies as inputs and outputs. (a) Integration with ASR. The dashed arrow indicates the original connection without VoiceFilter-Lite. (b) Neural network topology of the VoiceFilter-Lite model.}
     \label{fig:architecture}
     \vspace{-10pt}
\end{figure*}

\vspace{-3pt}
\section{VoiceFilter-Lite}

\subsection{Integration with ASR}

Our new VoiceFilter-Lite model focuses on improving ASR performance. Thus it is unnecessary to reconstruct any audio waveform from the masked spectrogram via inverse STFT. Instead, VoiceFilter-Lite operates as a frame-by-frame frontend signal processor that directly takes acoustic features as input, and outputs enhanced acoustic features. Here the acoustic features are exactly the same features being consumed by ASR models, such that the VoiceFilter-Lite model does not even need its own STFT operator. Since our ASR models take stacked log Mel-filterbank energies\footnote{We may simply use ``filterbanks'' to refer to log Mel-filterbank energies in this paper. We use $\log(1+x)$ to avoid negative values, where $1$ is a small value compared with the scale of the energies.} as input, our VoiceFilter-Lite model has three integration options: (1) Takes FFT magnitudes as input, and outputs enhanced FFT magnitudes, which will be used to compute filterbank energies; (2) Takes filterbank energies as input, and outputs enhanced filterbank energies; (3) Directly takes stacked filterbank energies as input. For example, the last option is demonstrated in Fig.~\ref{fig:VoiceFilterLite_IO}. One benefit of this design is that the VoiceFilter-Lite model is independent from the ASR model, such that it can be easily bypassed when the d-vector is not available from the user, unlike systems such as~\cite{denisov2019end}.

\subsection{Model topology}

In Fig.~\ref{fig:topology}, we show the neural network topology of the VoiceFilter-Lite model assuming we use stacked filterbank energies as the inputs. Although it looks similar to the previous work described in~\cite{wang2018voicefilter}, since VoiceFilter-Lite is designed for streaming ASR, we have made several changes to guarantee minimal latency: (1) We limit the convolutional layers to be 1D instead of 2D, meaning the convolutional kernels are for the frequency dimension only; (2) The LSTM layers must be uni-directional, and must be able to take streaming inputs. In practice, since frequency-only 1D-CNN is not as powerful as 2D-CNN, most of our models actually remove these CNN layers and purely consist of LSTM layers.

We assume the d-vector is available at runtime without additional computational cost. In a real application, users are usually prompted to follow an \emph{enrollment} process~\cite{enrollmentblog,wang2020version} before enabling speaker verification or voice filtering. During this one-off enrollment process, the d-vector is computed from the target user's recordings, and stored on the user's device.

At training time, the noisy audio is generated by mixing interference audio sources with the waveform of the clean audio, which is similar to~\cite{wang2018voicefilter}. However, in this work, to make our VoiceFilter-Lite model robust to various noise conditions, we add more variations to the noisification process: (1) the interference audio sources can be either speech from other speakers, or non-speech noise such as ambient noise or background music  --- we can also train the VoiceFilter-Lite model to predict the type of noise, which will be discussed in Section~\ref{sec:adapt_weight}; (2) the noises can be applied through either additive or reverberant operations~\cite{Kim2017}; (3) the signal-to-noise ratio (SNR) is also random within $1$dB to $10$dB.

\subsection{Asymmetric loss}
\label{sec:asym_loss}

Modern ASR models such as~\cite{he2019streaming} are usually trained with extensively augmented data, such as multistyle training (MTR)~\cite{lippmann1987multi,ko2017study,Kim2017} and SpecAugment~\cite{park2019specaugment}. Such ASR models already have great robustness against noise. If we add a voice filtering component to an existing ASR system, we must guarantee that the ASR performance does not degrade for any noise condition, which is a very challenging task. In fact, our early experiments show that when non-speech noise is present, the ASR performance could significantly degrade when we enable the voice filtering model. We observed that most of the degradation in Word Error Rate (WER) is from false deletions, which indicates significant \textbf{over-suppression} by the voice filtering model.

To overcome the over-suppression problem, we propose a new loss function for masking-based speech separation or enhancement, named \textbf{asymmetric loss}. Let $S_\mathrm{cln}(t,f)$ and $S_\mathrm{enh}(t,f)$ denote the clean and enhanced time-frequency spectrogram related features\footnote{Here the spectrogram can be either FFT magnitudes, filterbank energies, or stacked filterbank energies. The FFT magnitudes can also be power-law compressed like in~\cite{wilson2018exploring}.}, respectively. A conventional L2 loss is defined as:
\begin{equation}
    L = \sum_t \sum_f \Big( S_\mathrm{cln}(t,f) - S_\mathrm{enh}(t,f) \Big)^2
    .
    \vspace{-5pt}
\end{equation}

We would like to be more tolerant of under-suppression errors, and less tolerant of over-suppression errors. Thus, we define an asymmetric penalty function $g_\mathrm{asym}$ with penalty factor $\alpha > 1$:
\begin{equation}
    g_\mathrm{asym}(x, \alpha)=
\begin{cases}
x & \text{ if $x \leqslant 0$;}\\
\alpha \cdot x & \text{ if $x > 0$.}
\end{cases}
\end{equation}

Then the asymmetric L2 loss function can be defined as:
\begin{equation}
    L_\mathrm{asym} = \sum_t \sum_f \Big( g_\mathrm{asym} \big( S_\mathrm{cln}(t,f) - S_\mathrm{enh}(t,f) , \alpha \big) \Big)^2
    .
    \vspace{-5pt}
\end{equation}

\subsection{Adaptive suppression strength}
\label{sec:adapt_weight}

As mentioned before, modern ASR models are usually already robust against non-speech noise. Having a voice filtering model that performs an additional step of feature masking will often harm the ASR performance. Even with asymmetric loss, while our model significantly improves ASR performance under speech noise, it can still degrade ASR performance under non-speech noise.

One way to mitigate the performance degradation is to have an additional compensation to the over-suppression at inference time. Let $S_\mathrm{in}^{(t)}$ and $S_\mathrm{enh}^{(t)}$ denote the input and enhanced spectrogram related features at time $t$, respectively. The final compensated output would be:
\begin{equation}
    \label{eq:compensate}
    S_\mathrm{out}^{(t)} = w \cdot S_\mathrm{enh}^{(t)} + (1-w) \cdot S_\mathrm{in}^{(t)}
    .
\end{equation}
Here $w$ is the suppression strength. When $w=0$, voice filtering is completely disabled; and when $w=1$, there is no compensation.

In practice, instead of using a fixed value of $w$, we wish to use a larger $w^{(t)}$ when the voice filtering model improves ASR, and a smaller $w^{(t)}$ when it hurts ASR. Thus, we add a second binary classification output to the model, which predicts whether a feature frame is from overlapped speech (class label $1$) or not (class label $0$) as shown in Fig.~\ref{fig:topology}. If we denote the noise type prediction as $f_\mathrm{adapt}(S_\mathrm{in}^{(t)}) \in [0,1]$, the adaptive suppression strength at time $t$ can be defined as
\begin{equation}
    w^{(t)} = \beta \cdot w^{(t-1)} + (1-\beta) \cdot \big( a \cdot f_\mathrm{adapt}(S_\mathrm{in}^{(t)}) + b \big) ,
\end{equation}
where $a>0$ and $b \geqslant 0$ define a linear transform, and $0 \leqslant \beta < 1$ is a moving average coefficient to make the suppression more smooth.

\begin{table*}[t]
\centering
  \caption{WER (\%) for VoiceFilter-Lite models. ASR is trained and evaluated on LibriSpeech.}
  \vspace{-5pt}
  \label{tab:wer_librispeech}
  \begin{tabular}{| c | c | c | c | c | c | c | c | c |}
    \hline
    \bf \multirow{2}{*}{Feature} & \bf \multirow{2}{*}{Loss} & \bf Suppression & \bf \multirow{2}{*}{Clean} & \multicolumn{2}{|c|}{\bf Non-speech noise} & \multicolumn{2}{|c|}{\bf Speech noise} & \bf \multirow{2}{*}{Size}\\ \cline{5-8}
    && \bf strength && \bf Additive & \bf Reverb & \bf Additive & \bf Reverb & \\ \hline
    \multicolumn{3}{|c|}{No voice filtering} & 8.6 & 35.7 & 58.5 & 77.9 & 79.3 & N/A \\ \hline
    \multirow{2}{*}{FFT magnitude} & L2 & $w=1.0$ & 9.1 & 21.5 & 48.3 & 25.5 & 54.2 & \multirow{2}{*}{6.8 MB} \\ \cline{2-8}
    & asym L2, $\alpha=10$ & $w=1.0$ & 8.8 & 24.1 & 50.8 & 35.5 & 60.6 & \\ \cline{3-8} \hline
    \multirow{2}{*}{Filterbank} & L2 & $w=1.0$ & 9.3 & 23.4 & 48.9 & 25.4 & 55.6 & \multirow{2}{*}{5.8 MB} \\ \cline{2-8}
    & asym L2, $\alpha=10$ & $w=1.0$ & 8.6 & 24.8 & 49.8 & 30.6 & 58.4 &  \\ \hline
    \multirow{3}{*}{Stacked filterbank} & L2 & $w=1.0$ & 8.9 & 22.2 & 48.2 & 23.5 & 53.7 & \multirow{3}{*}{6.8 MB} \\ \cline{2-8}
    & \multirow{2}{*}{asym L2, $\alpha=10$} & $w=1.0$ & 8.8 & 23.9 & 49.7 & 30.6 & 57.8 &  \\ \cline{3-8}
        && $w=0.6$ & 8.6 & 24.4 & 50.7 & 42.0 & 60.2 & \\ \cline{3-8} \hline
  \end{tabular}
  \vspace{-5pt}
\end{table*}

\begin{table*}[t]
\centering
  \caption{WER (\%) for VoiceFilter-Lite models. ASR is trained on multi-domain data, and  evaluated on vendor-collected speech queries.}
  \vspace{-5pt}
  \label{tab:wer_real}
  \begin{tabular}{| c | c | c | c | c | c | c | c | c |}
    \hline
    \bf \multirow{2}{*}{Feature} & \bf \multirow{2}{*}{Loss} & \bf Suppression & \bf \multirow{2}{*}{Clean} & \multicolumn{2}{|c|}{\bf Non-speech noise} & \multicolumn{2}{|c|}{\bf Speech noise} & \bf \multirow{2}{*}{Size}\\ \cline{5-8}
    && \bf strength && \bf Additive & \bf Reverb & \bf Additive & \bf Reverb & \\ \hline
    \multicolumn{3}{|c|}{No voice filtering} & 15.2 & 21.1 & 29.1 & 56.5 & 53.8 & N/A \\ \hline
    \multirow{2}{*}{FFT magnitude} & L2 & $w=1.0$ & 15.4 & 27.0 & 36.9 & 25.1 & 36.8 & \multirow{2}{*}{6.8 MB} \\ \cline{2-8}
    & asym L2, $\alpha=10$ & $w=1.0$ & 15.2 & 22.6 & 31.2 & 32.0 & 37.8 & \\ \cline{3-8} \hline
    \multirow{2}{*}{Filterbank} & L2 & $w=1.0$ & 15.3 & 28.5 & 38.3 & 26.5 & 38.5 & \multirow{2}{*}{5.8 MB} \\ \cline{2-8}
    & asym L2, $\alpha=10$ & $w=1.0$ & 15.3 & 26.6 & 35.6 & 27.5 & 37.4 &  \\ \hline
    \multirow{5}{*}{Stacked filterbank} & L2 & $w=1.0$ & 16.7 & 26.8 & 36.2 & 26.8 & 37.4 & \multirow{4}{*}{6.8 MB} \\ \cline{2-8}
    & \multirow{4}{*}{asym L2, $\alpha=10$} & $w=1.0$ & 15.8 & 25.7 & 34.4 & 27.4 & 36.7 &  \\ \cline{3-8}
        && $w=0.3$ & 15.2 & 21.7 & 29.6 & 42.1 & 43.6 & \\ \cline{3-8}
    && \multirow{2}{*}{Adaptive $w^{(t)}$} & 15.3 & 21.3 & 29.3 & 28.8 & 37.2 & \\ \cline{4-9}
    & & & 15.4 & 21.1 & 29.0 & 31.4 & 39.1 & 2.2 MB \\ \hline
  \end{tabular}
  \vspace{-15pt}
\end{table*}

\subsection{Model quantization}
\label{sec:quantization}

Raw TensorFlow graphs store network parameters in 32-bit floating point values, and are not well optimized for on-device inference. We quantize the network parameters to 8-bit integers with dynamic range quantization~\cite{alvarez2016efficient,shangguan2019optimizing}, and serialize the models to the FlatBuffer TensorFlow Lite format. This helps us significantly reduce memory cost, and make better use of the optimized hardware instructions for integer arithmetic.

\section{Experiments}
\vspace{-5pt}
\subsection{Metrics}

Our VoiceFilter-Lite model focuses on improving on-device ASR, thus we use the Word Error Rate (WER) as our metric. Specifically, we care about the WER before and after applying VoiceFilter-Lite under various noise conditions. Since the VoiceFilter-Lite model does not reconstruct any waveform from the enhanced acoustic features, and given the focus on speech recognition, we are not going to report metrics that have been widely used for conventional source separation, such as signal-to-noise ratio (SNR) or source-to-distortion ratio (SDR)~\cite{vincent2006performance,le2019sdr}.

\subsection{Model and data}

In our experiments, the VoiceFilter-Lite model has 3 LSTM layers, each with 512 nodes, and a final fully connected layer with sigmoid activation. If noise type prediction is enabled, it has another two feedforward layers, each with 64 nodes, and is trained with a hinge loss~\cite{rosasco2004loss} for binary classification.
The training data consist of: (1) The LibriSpeech training set; and (2) a vendor-collected dataset of realistic speech queries. The d-vectors are generated with the same speaker encoder LSTM that has been used by~\cite{jia2019direct} and~\cite{ding2019personal}.

For the WER evaluations, we use different testing sets for different ASR models, as described in Section~\ref{sec:result_librispeech} and Section~\ref{sec:result_real}, respectively. For all evaluations, we apply different noise sources and room configurations to the data. We use ``Clean" to denote the original non-noisified data, although they could be quite noisy already. Our non-speech noise source consists of ambient noises recorded in cafes, cars, and quiet environments, as well as music or other types of sounds; the speech noise source is a distinct development set without overlapping speakers from the testing set. We evaluate with two types of room conditions:  ``Additive" means directly adding the noise waveform to the clean waveform; ``Reverb" consists of 3 million convolutional room impulse responses (RIR) generated by a room simulator~\cite{Kim2017}. When applying the noises, the SNR is drawn from a uniform distribution from $1$dB to $10$dB for both additive and reverberant conditions.

\subsection{Results on LibriSpeech}
\label{sec:result_librispeech}

First, we use an RNN-transducer~\cite{graves2012sequence} based ASR model trained on the LibriSpeech training set~\cite{he2019streaming}, and compute the WER on LibriSpeech as well (``test-clean'' and ``test-other''). The results are reported in Table~\ref{tab:wer_librispeech}. Without voice filtering, the ASR model has a WER of 8.6\% on the LibriSpeech testing set (5.2\% on the ``test-clean'' subset). But after applying noises to the testing data, the WER becomes much worse.

With a VoiceFilter-Lite model, no matter which feature we use as input, the WER on the noisy conditions can be significantly reduced, but often hurting the WER in the clean condition. By training with asymmetric L2 loss, or applying a fixed suppression strength, the WER degradation in the clean case could be mitigated, at the cost of losing some performance gains in noisy conditions. For example, the filterbank-based model provides 47.3\% (rel. 60.7\%) WER improvements under additive speech noise conditions, without hurting WER in the clean case.

\begin{figure}
	\centering
	\includegraphics[width=0.48\textwidth]{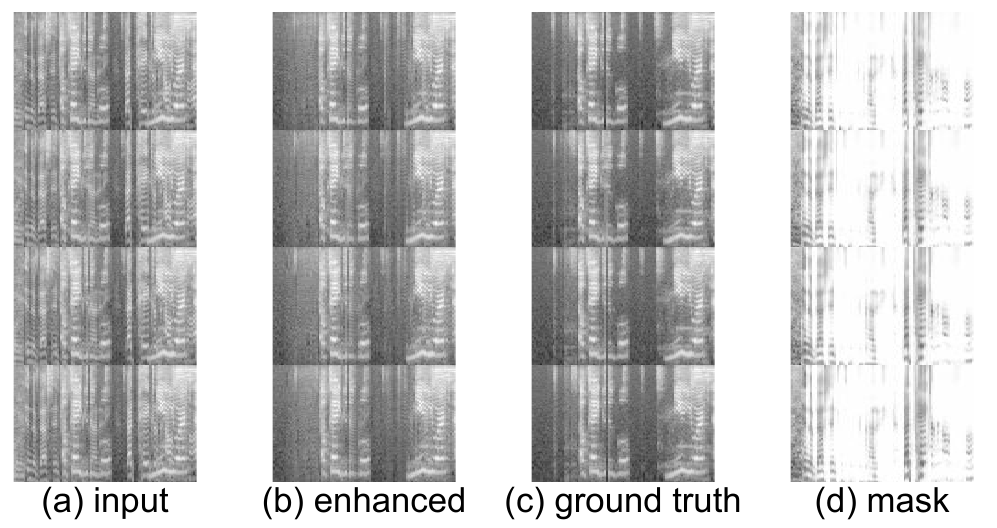}
	\vspace{-10pt}
	\caption{Example stacked filterbanks from a segment of 3 seconds during training. x-axis is time and y-axis is filterbank.}
	\label{fig:spectrogram}
	\vspace{-15pt}
\end{figure}

\subsection{Results on realistic speech queries}
\label{sec:result_real}

Although the above results demonstrated the success of VoiceFilter-Lite on LibriSpeech, we would also like to evaluate on more realistic data. We trained another on-device streaming RNN-transducer ASR model~\cite{he2019streaming} using data from multiple domains including YouTube and anonymized voice search, then evaluate the WER on a vendor-collected dataset of realistic speech queries. This dataset was collected via a Web-based tool, and consists of about 20 thousand utterances from 230 speakers. The task here is much more challenging than LibriSpeech because: (1) There are more variations of prosody, sentiment, accent, and acoustic conditions in real speech queries than read speech; (2) We are evaluating across the domain, since the type of evaluation data is not represented in ASR training.

From the results shown in Table~\ref{tab:wer_real}, we can see that, if the VoiceFilter-Lite model is trained with L2 loss, although the WERs on speech noise are greatly reduced, the WERs on non-speech noise largely degrade. This observation is consistent no matter what features we use for VoiceFilter-Lite.
A model like this could not be used in production, since it harms performance in certain conditions. However, with the asymmetric L2 loss introduced in Section~\ref{sec:asym_loss}, the degradation on non-speech noise is much smaller, although we also see less improvement on speech noise at the same time.

For models using stacked filterbanks as features, we also experimented with suppression compensation. If we use a fixed strength $w=0.3$, the degradation on non-speech noise is further reduced, but the improvement on speech noise is also compromised. If we use adaptive suppression strength with a moving average coefficient $\beta=0.8$ (2nd last row in Table~\ref{tab:wer_real}), we observe almost no WER degradation on clean and non-speech noise conditions, and still a large improvement on speech noise conditions  --- 27.7\% (rel. 49.0\%) WER improvement under additive speech noise condition. This meets our goal of having an \textbf{always harmless and sometimes helpful} model that is safe to use in real applications. By reducing each LSTM layer to only 256 nodes, the model size becomes 2.2 MB (last row in Table~\ref{tab:wer_real}), but still with similar WER performance.  

\vspace{-5pt}
\subsection{Discussions}
\vspace{-5pt}
For our three different features, \emph{i.e.} the 513-dim FFT magnitudes, the 128-dim log Mel-filterbank energies, and the 512-dim stacked-by-4 log Mel-filterbank energies, our results are quite similar. However, we argue that using stacked filterbanks as model inputs and outputs is largely preferred: (1) First, the stacked filterbanks contain more contextual information than single-frame FFT or filtebanks; (2) Second, since frame subsampling is usually applied to reduce redundant information from frame stacking, the network runs less often when taking stacked filterbanks as input, which largely reduces CPU cost with the same number of network parameters. Examples of stacked filterbanks during training are visualized in Fig.~\ref{fig:spectrogram}.

\vspace{-5pt}
\section{Conclusions}
\vspace{-5pt}
In this paper, we described VoiceFilter-Lite, a tiny and fast model that performs targeted voice separation in a streaming fashion, as part of an on-device ASR system. Since modern ASR models are already trained with a diverse range of noise conditions, we need to guarantee that the voice separation model does not hurt the ASR performance, especially under non-speech noise, and reverberant room conditions. We achieved this by training our model with an asymmetric loss function, and applying an adaptive suppression strength at runtime. Combining these novel efforts, we developed a 2.2 MB model that has no WER degradation on the clean and non-speech noise conditions we measured, while largely improving ASR performance on overlapped speech.

\vspace{-5pt}
\section{Acknowledgements}
\vspace{-5pt}
The authors would like to thank Philip Chao, Sinan Akay, John Han, Stephen Wu, Yiteng Huang, Jaclyn Konzelmann and Nino Tasca for the support and helpful discussions. 

\bibliographystyle{IEEEtran}
\bibliography{mybib}

\end{document}